\documentclass[conference,a4paper]{IEEEtran}
\IEEEoverridecommandlockouts

\usepackage{fancyhdr}
\usepackage[dvips]{graphicx}
\usepackage{indent}
\usepackage{amsmath,amssymb}
\usepackage{paralist}
\usepackage{eclbkbox}
\usepackage{subfigure}
\usepackage{multirow}
\usepackage{bm}

\makeatletter
\def\theequation{\arabic{equation}}
\def\tbcaption{\def\@captype{table}\caption}
\def\figcaption{\def\@captype{figure}\caption}
\makeatother

{\pointlessenum\begin{enumerate}}%
{\end{enumerate}}

\begin{document}
\title{An Adaptive Learning Method of Personality Trait Based Mood in Mental State Transition Network\\by Recurrent Neural Network
\thanks{\copyright 2014 IEEE. Personal use of this material is permitted. Permission from IEEE must be obtained for all other uses, in any current or future media, including reprinting/republishing this material for advertising or promotional purposes, creating new collective works, for resale or redistribution to servers or lists, or reuse of any copyrighted component of this work in other works.}}

\author{\IEEEauthorblockN{Takumi Ichimura}
\IEEEauthorblockA{Department of Management and Systems,\\ Prefecture University of Hiroshima,\\ Hiroshima, 734-8558 Japan\\
E-mail: ichimura@pu-hiroshima.ac.jp}
\and
\IEEEauthorblockN{Kosuke Tanabe}
\IEEEauthorblockA{NEC Solution Innovators, Ltd.\\
Email: bakabonn009@gmail.com}
\and
\IEEEauthorblockN{Toshiyuki Yamashita}
\IEEEauthorblockA{Department of Human Sciences,\\ Graduate School of Humanities,\\ Tokyo Metropolitan University,\\ Tokyo, 192-0397 Japan\\ 
Email: yamashita-toshiyuki@tmu.ac.jp}
}

\maketitle

\pagestyle{fancy}{
\fancyhf{}
\fancyfoot[R]{}}
\renewcommand{\headrulewidth}{0pt}
\renewcommand{\footrulewidth}{0pt}

\begin{abstract}
Mental State Transition Network (MSTN) is a basic concept of approximating to human psychological and mental responses. A stimulus calculated by Emotion Generating Calculations (EGC) method can cause the transition of mood  from an emotional state to others. In this paper, the agent can interact with human to realize smooth communication by an adaptive learning method of the user's personality trait based mood. The learning method consists of the profit sharing (PS) method and the recurrent neural network (RNN). An emotion for sensor inputs to MSTN is calculated by EGC and the variance of emotion leads to the change of mental state, and then the sequence of states forms an episode. In order to learn the tendency of personality trait effectively, the ineffective rules should be removed from the episode. PS method finds out a detour in episode and should be deleted. Furthermore, RNN works to realize the variance of user's mood. Some experimental results were shown the success of representing a various human's delicate emotion.
\end{abstract}


\IEEEpeerreviewmaketitle

\section{Introduction}
\label{sec:Introduction}
Our research group proposed a method to calculate the agent's emotion from the contents of utterances and to express emotions which are aroused in computer agent\cite{Mera2002, Mera2010, Ichimura2013}. Emotion Generating Calculations (EGC) method \cite{Mera2002} based on the Emotion Eliciting Condition Theory \cite{Elliott92} can decide whether an event arouses pleasure or not and quantify the degree under the event. Calculated emotions effect a change pf the mood to the agent. Ren \cite{Ren06} describes Mental State Transition Network (MSTN) which is the basic concept of approximating to human psychological and mental responses. The assumption of discrete emotion state is that human emotions are classified into some kinds of stable discrete states, called ``mental state,'' and the variance of emotions occurs in the transition from a state to other state with an arbitrary probability. Mera et al. \cite{Mera2010} developed a computer agent which can transit a mental state in MSTN based on analysis of emotion by EGC method. EGC calculates the type and the degree of the aroused emotion to transit mental state \cite{Mera2010}.

This paper describes a method to learn a general tendency to change the user's mental state by Profit sharing (PS) method \cite{Grefenstette88, Miyazaki94, Sutton98} and recurrent neural networks \cite{Grau2013}.

 An emotion for sensor inputs to MSTN is calculated by EGC and the variance of emotion leads to the change of mental state, and then the sequence of states forms an episode. In order to learn the tendency of personality trait effectively, the ineffective rules should be removed from the episode. Profit sharing (PS) method \cite{Grefenstette88, Miyazaki94} is used to find out a detour in episode and to remove it. Reward to the episode is an emotion value which is calculated by EGC method. Pleasure / displeasure and its intensity is calculated from the final action in episode. Furthermore, to pay attention to personal difference and to realize it, Back Propagation through Time (BPTT) learning algorithm of recurrent neural networks (RNN) is used to learn the temporal difference in MSTN.

We analyzed the transition probability in trained MSTN to discuss the relation between the state transition and the Big Five Makers (BFM) \cite{Goldberg1992}, which was designed to assess the constellation of traits defined by the Five Factor Theory of Personality such as `Openness', `Conscientiousness', `Extraversion', `Agreeableness', and `Neuroticism'. In order to verify the effectiveness of the learning method, some experimental results by applying the movie scenario were shown in this paper.

\section{Mental State Transition Learning Network}
\label{sec:MentalStateTransitionLarningNetwork}
\subsection{Mental State Transition Network} 
\begin{figure}[btp]
\begin{center}
\includegraphics[scale=0.5]{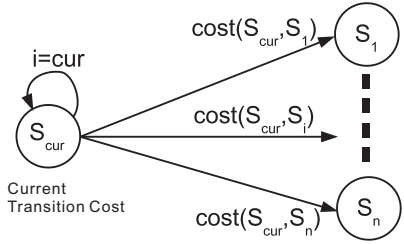}
\caption{Transition Cost}
\label{fig:TransitionCost}
\end{center}
\end{figure}

\begin{figure}[btp]
\begin{center}
\includegraphics[scale=0.5]{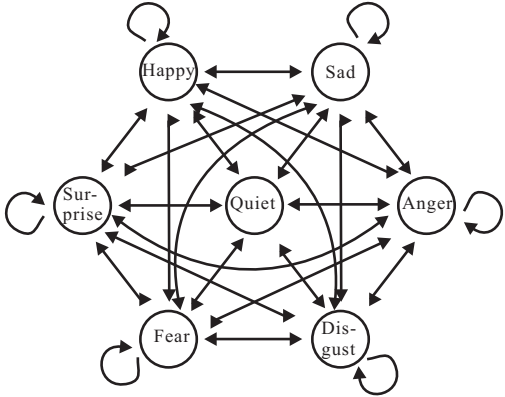}
\caption{Concept of MSTN}
\label{fig:MSTN}
\end{center}
\end{figure}

MSTN, proposed by Ren \cite{Ren06}, represents the basic concept of approximating to human physiological and mental responses. He focuses not only information included in the elements of phonation, facial expressions, and speech, but also human psychological characteristics based on the latest achievements of brain science and psychology in order to derive transition networks for human psychological states. The assumption of discrete emotion state is that human emotions are classified into some kinds of stable discrete states, called ``mental state'', and the variance of emotions occurs in the transition from a state to other state with a probability. The probability of transition is called ``transition cost'' and it has different values among transitions. Moreover, with no stimulus from the external world, the probability may converge to fall into a certain value as if the confusion of the mind leaves and is relieved. On the contrary, with a stimulus from external world and/or attractive thought in internal world, the continuous accumulated emotional energy cannot jump to the next mental state and remains in its mental state still. The simulated model of mental state transition network \cite{Ren06} describes the simple relations among some kinds of stable emotions and the corresponding transition probability. The probability was calculated from analysis of many statistical questionnaire data.
As shown in Fig.\ref{fig:MSTN}, the MSTN denotes a mental state as a node, a set of some kinds of mental state $\mathcal{S}$, the current emotion state $\mathcal{S}_{cur}$ , and the transition cost $cost(\mathcal{S}_{cur}, \mathcal{S}_{i})$, which is the transition cost as shown in Fig.\ref{fig:TransitionCost}.

In \cite{Ren06}, six kinds of mental states and a quiet state are considered for questionnaire. That is, the transition table of $cost(\mathcal{S}_{i}, \mathcal{S}_{j})$, $i=1,2,\cdots,7$, $j=1,2,\cdots,7$ is prepared. The experiment for participants was examined without stimulus from external world. Each participant fills in the numerical value from 1 to 10 that means the strength of relation among mental states. Moreover, the same questionnaire was examined under the condition with the stimulus from external world. The 200 participants answered the questionnaire. The numerical values in Table \ref{tab:TransitionCost} show the statistical analysis results. The transition cost from each current state to the next state is summarized to $1.0$.

\begin{table*}[htbp]
\begin{center}
\caption{Transition Cost in MSTN}
\begin{tabular}{l|c|ccccccc}
\hline
\multicolumn{2}{c|}{ }      & \multicolumn{7}{c}{next mental state}\\
\cline{3-9}
\multicolumn{2}{c|}{ }      & happy & quiet & sad   & surprise & angry & fear & disgust\\ \hline
&happy    & 0.421 & 0.362 & 0.061 & 0.060 & 0.027 & 0.034 & 0.032\\
&quiet    & 0.213 & 0.509 & 0.090 & 0.055 & 0.039 & 0.051 & 0.042\\
current&sad      & 0.084 & 0.296 & 0.320 & 0.058 & 0.108 & 0.064 & 0.068\\
mental &surprise & 0.190 & 0.264 & 0.091 & 0.243 & 0.086 & 0.076 & 0.048\\
state  &angry    & 0.056 & 0.262 & 0.123 & 0.075 & 0.293 & 0.069 & 0.121\\
&fear     & 0.050 & 0.244 & 0.137 & 0.101 & 0.096 & 0.279 & 0.092\\
&disgust  & 0.047 & 0.252 & 0.092 & 0.056 & 0.164 & 0.075 & 0.313\\ \hline
\end{tabular}
\label{tab:TransitionCost}
\end{center}
\end{table*}

\subsection{EGC with MSTN}
Even if there are not any signals from external world, the mental state will change small. In this case, the transition costs represented in Table \ref{tab:TransitionCost} are adopted to calculate by using EGC.
In this paper, we assume that the stimulus from external world is the utterance of the user and the transition cost is calculated as follows.

\begin{equation}
cost(\mathcal{S}_{i}, \mathcal{S}_{j})=1-\frac{\#(\mathcal{S}_{i} \rightarrow \mathcal{S}_{j})}{\sum_{j=1}^{7} \#(\mathcal{S}_{i} \rightarrow \mathcal{S}_{j})},
\label{eq:TransitionCost}
\end{equation}
where $\#(\mathcal{S}_{i} \rightarrow \mathcal{S}_{j})$ is the number of transition from mental state $\mathcal{S}_{i}$, $1 \leq i \leq 7$ to $\mathcal{S}_{j}$, $1 \leq j \leq 7$. The transition cost is calculated by using the total of $\#(\mathcal{S}_{i} \rightarrow \mathcal{S}_{j})$ for all mental state. If the transition cost is high in Eq.(\ref{eq:TransitionCost}), the transition is difficult to happen.

\begin{figure}[hbtp]
\begin{center}
\includegraphics[scale=0.3]{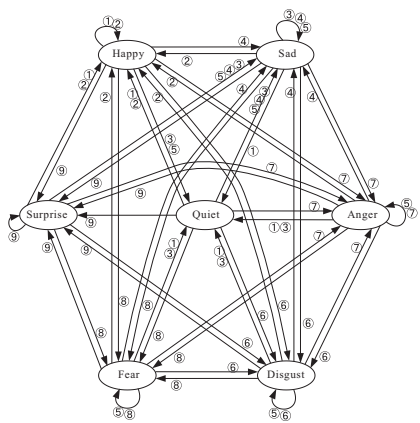}
\caption{MSTN with EGC}
\label{fig:MSTN_EGC}
\end{center}
\end{figure}

Eq.(\ref{eq:TransitionCost_next}) calculates the next mental state from the current mental state $\mathcal{S}_{cur} \in \mathbf{S}$ by using the emotion vector.
\begin{equation}
next=\arg \max_{k} \frac{e_{k}}{cost(\mathcal{S}_{cur}, \mathcal{S}_{i})}, \ 1 \leq k \leq 9
\label{eq:TransitionCost_next}
\end{equation}
The emotion vector consists of 9 kinds of emotion groups which are classified 28 kinds of emotions as shown in Table \ref{tab:classofemotion}. Fig.\ref{fig:MSTN_EGC} shows the MSTN by using EGC. The circled numbers in Fig.\ref{fig:MSTN_EGC} are the number in the left side of Table \ref{tab:classofemotion}. The $e_{k}$ $(1 \leq k \leq 9)$ shows the strength of emotion group $k$ and takes the maximum value of elements belonged in each set $e_{k}$ as follows.

\begin{description}
\item $e_{1}=\max (e_{gloating}, e_{hope}, \cdots, e_{shy})$
\item $e_{2}=\max (e_{joy}, e_{happy\_for})$
\item $\vdots$
\item $e_{9}=\max (e_{surprise})$

\end{description}
\begin{table}[tbp]
\begin{center}
\caption{Classification of Generated Emotion}
\begin{tabular}{c|c}
\hline
No. & Emotion \\ \hline
  & gloating, hope, satisfaction, relief, pride,\\
1 & admiration, liking, gratitude, gratification,\\
  & love, shy \\ \hline
2 & joy,  happy\_for \\ \hline
3 & sorry-for, shame, remorse \\ \hline
4 & fear-confirmed, disappointment, sadness \\ \hline
5 & distress, perplexity \\ \hline
6 & disliking, hate \\ \hline
7 & resentment, reproach, anger \\ \hline
8 & fear \\ \hline
9 & surprise \\ \hline
\end{tabular}
\label{tab:classofemotion}
\end{center}
\end{table}

The $emo$ in Eq.(\ref{eq:selectemotion}) calculates the maximum emotion group according to the transition cost between current state and next state.
\begin{equation}
emo_{k}=\arg \max_{k} \frac{e_{k}}{cost(\mathcal{S}_{cur}, next(\mathcal{S}_{cur},k))}, \ 1 \leq k \leq 9,
\label{eq:selectemotion}
\end{equation}
where $next(\mathcal{S}_{cur}, k)$ is next mental state from the current state by selecting emotion group $k$.

\subsection{Mental State and Action}
\begin{figure}[tbp]
\begin{center}
\includegraphics[scale=0.5]{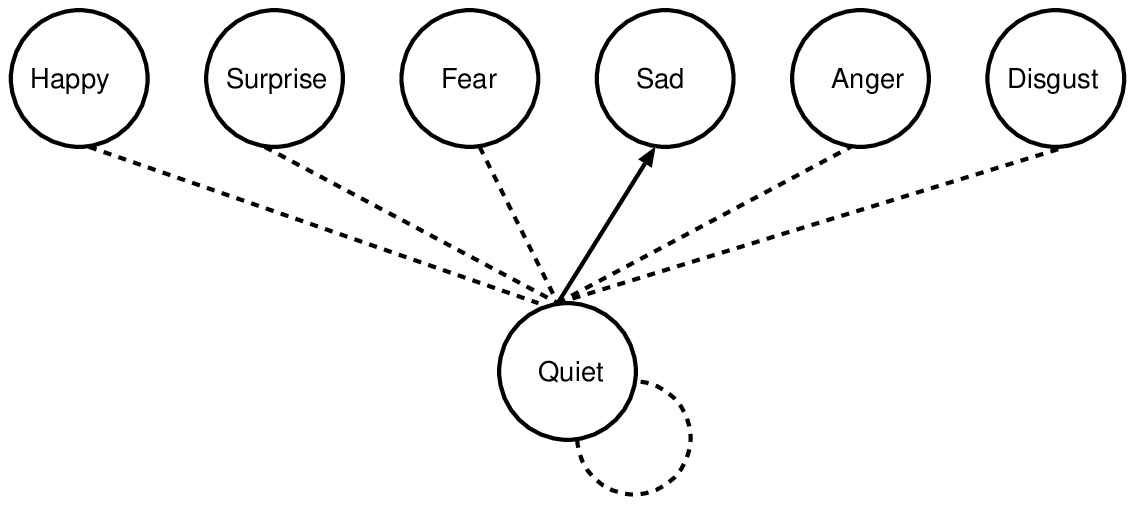}
\caption{Example of State and Action in MSTN}
\label{fig:stateaction_mentalstate_1}
\end{center}
\end{figure}

\begin{figure}[tbp]
\begin{center}
\includegraphics[scale=0.5]{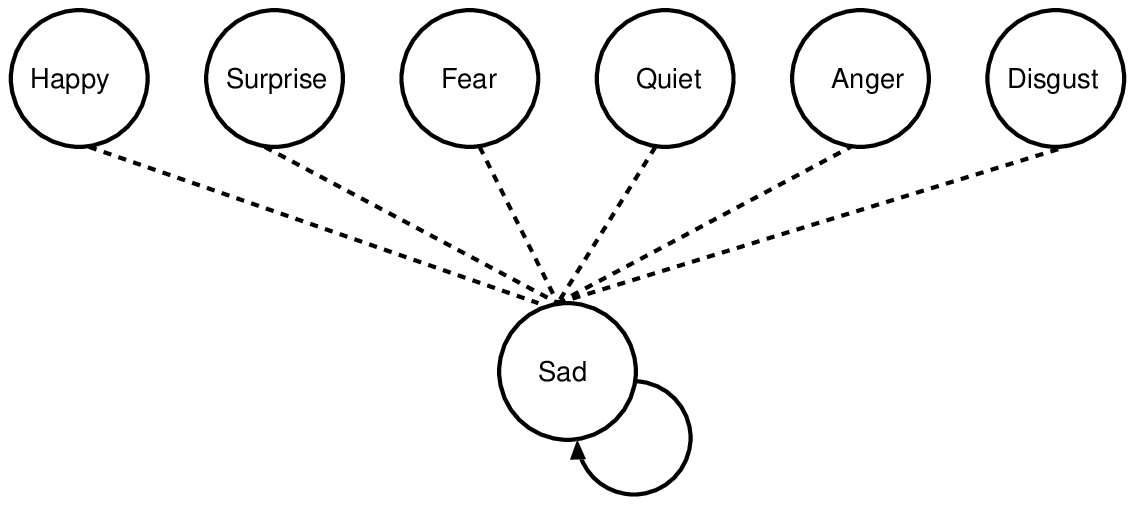}
\caption{Example of State and Action in MSTN}
\label{fig:stateaction_mentalstate_2}
\end{center}
\end{figure}

Fig.\ref{fig:stateaction_mentalstate_1} shows an example of when the current state is `quiet' and the next state is `sad.' Although the mental state `quiet' means the stable state representing usual mind without affections, the transition from this stable state to other states is occurred easy, if human feels some stimuli from external world. However, each state in MSTN often remains to be in a loop path to back the same state, because it is a tendency that human continues to feel same emotions, once human feels strong stimulus to reach the depth psyche. For example, Fig.\ref{fig:stateaction_mentalstate_2} shows the next state as shown in Fig.\ref{fig:stateaction_mentalstate_1}, that is, the mental state falls into the `sad' state and cannot break out of the loop. Once human falls into the specified mental state deeply, he/she traces the path to same mental state again until he/she perceives another stimulus by himself/herself in his/her surrounding environment. Moreover, human often disposes to psychological variability by a weak stimulus from external world. Even if MSTN receives such a stimulus, driving force for the movement toward next mental state is not occurred, because the signal calculated by EGC is a little. Therefore, mental state falls into some perturbation due to weak stimulus. Although such perturbation or the bemusement for him/her is not always waste in the loop, the events which the agent meets in MSTN apt to construct a same episode. The aim of agent in MSTN avoids such a situation and makes smooth communication. In this paper, the agent works to find the detours and to eliminate them. The reinforcement learning method is proposed to avoid such a situation and to make smooth communication described in the following section.

\section{Learning Personal Mental State}
\label{sec:LearningPersonalMentalState}
This section describes the algorithms of two kinds of Reinforcement Learning methods briefly. 

\subsection{Profit Sharing}
Multi agent systems have been developed in the field of Artificial Intelligence. Each agent is designed to work some schemes based on many rules which indicate knowledge of the agent world or relationship among the agents. However, the knowledge or relationship is not always effective to survive in their environment, because the agent will discard a partial of knowledge if its environment changes dynamically. Reinforcement Learning \cite{Sutton98} is known to be worth to realize the cooperative behavior among agents even if little knowledge is provided with initial condition. The multi agent system works to share a given reward among all agents.

Especially, PS method \cite{Grefenstette88}, \cite{Miyazaki94} is an effective exploitation of reinforcement learning to adapt to a given environment. In PS, an agent learns a policy based on the reward that is received from the environment when it reaches a goal state. It is important to design a reinforcement function that distributes the received reward to each action rule in the policy. In PS, the rule $r_{i}$ is $(s, a)$ for possible action $a$ to a given sensory input $x$ to $s$. The rule ``If $x$ then $a$.'' is also written by $\overrightarrow{xa}$. PS does not estimate the value function and computes weight of rules $S_{r_{i}}$ for $(s, a)$. The episode is determined from the start state to the terminal state which the agent achieves the goal at time $i$ and then a reward ${\bf R}$ is provided. The PS gives the partial reward of ${\bf R}$ to the fired rule $(s_{i}, a_{i})$ in an episode($i<W$). The partial ${\bf R}$ is determined by the value function $f(i, {\bf R}, W)$. Each rule is reinforced by the sum of current weight and slanted reward. That is,
\begin{equation}
S_{r_{i}}=S_{r_{i}}+f_{i}, \: i=0, 1, \cdots, W-1,
\label{eq:profitsharing-1}
\end{equation}
where $S_{r_{i}}$ means the weight of the $i$th rule of an episode, $f_{i}$ is the reinforce function and means the reinforce value at the $-i$ step from obtaining ${\bf R}$.

\begin{figure}[btp]
\begin{center}
\includegraphics[scale=0.7]{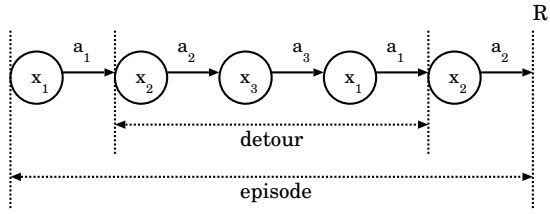}
\caption{The episode and the detour}
\label{fig:detour}
\end{center}
\end{figure}

The detour as shown in Fig.\ref{fig:detour} is the sequence of rules when the difference rules are selected for the same sensory input. There is a detour $(\overrightarrow{x_{2}a_{2}}, \overrightarrow{x_{3}a_{3}}, \overrightarrow{x_{1}a_{1}})$ in the sequence $(\overrightarrow{x_{1}a_{1}}, \overrightarrow{x_{2}a_{2}}, \overrightarrow{x_{3}a_{3}}, \overrightarrow{x_{1}a_{1}}, \overrightarrow{x_{2}a_{2}})$ in Fig.\ref{fig:detour}. The rules in the detour may occur some ineffective rules. The ineffective rule is always on the detour from the episode. The other rules are called the effective rule. If the competition between ineffective rules and effective rules exists, the ineffective rules are not reinforced. If the reinforcement function satisfies the ineffective rule suppression theorem, the reinforcement function is able to distribute more reward to effective rules than ineffective ones. In order to suppress such ineffective rules, the forgettable PS method is proposed. 

\begin{equation}
L \sum_{j=1}^{w} f_{j} < f_{i-1}, \forall i= 1, 2, \cdots, W,
\label{eq:profitsharing-2}
\end{equation}
where $W$ is the maximum length of episode and $L$ is the maximum number of effective rules. The reinforcement function decreases in a geometric series in the following.
\begin{eqnarray}
f_{i}=\frac{1}{M} f_{i-1}, i=1, 2, \cdots, W-1,
\label{eq:profitsharing-3}
\end{eqnarray}
where $M(\geq L+1)$ is a discount rate.
Eq.(\ref{eq:profitsharing-3}) reinforces the rule from $i=1$ to $i=W$ in an episode.

The algorithm of PS is as follows.
\begin{center}
%\begin{indentation}{0.1zw}{0.1zw}
\begin{indentation}{0.1cm}{0.1cm}
\begin{breakbox}
\smallskip
\begin{enumerate}[Step 1)]
\item Initialize $S_{r_{i}}$ arbitrarily.
\item Repeat (for each episode):
\begin{enumerate}
\item Initialize $r_{i}$ and $W$.
\item Repeat (for each step of episode):
\begin{enumerate}
\item $a \leftarrow$ action given by $\pi$ for $\mathcal{S}$ at state $x$ 
\item Take action $a$; observe reward, ${\bf R}$ and next state $\acute{x}$ 
\item $\forall i, i \leftarrow i+1$, set $r_{0}=\vec{x}_{a}$
\item If $R \neq 0$, set $f_{0}={\bf R}$ and calculate the following.
\begin{equation}
S_{r_{i}}=S_{r_{i}}+f_{i}, i=0, 1, \cdots, W-1,
\end{equation}
where $f_{i}=\frac{1}{M}f_{i-1}, i= 1, 2, \cdots, W-1$.
\item $x \leftarrow \acute{x}, W=W+1$
\end{enumerate}
\item until $x$ is terminal  
\end{enumerate}
\smallskip
\end{enumerate}
\end{breakbox}
\end{indentation}
\figcaption{The algorithm of PS}
\label{fig:algorithm-profitsharing}
\end{center}

\subsection{Recurrent Neural Network}
\begin{figure}[btp]
\begin{center}
\includegraphics[scale=0.35]{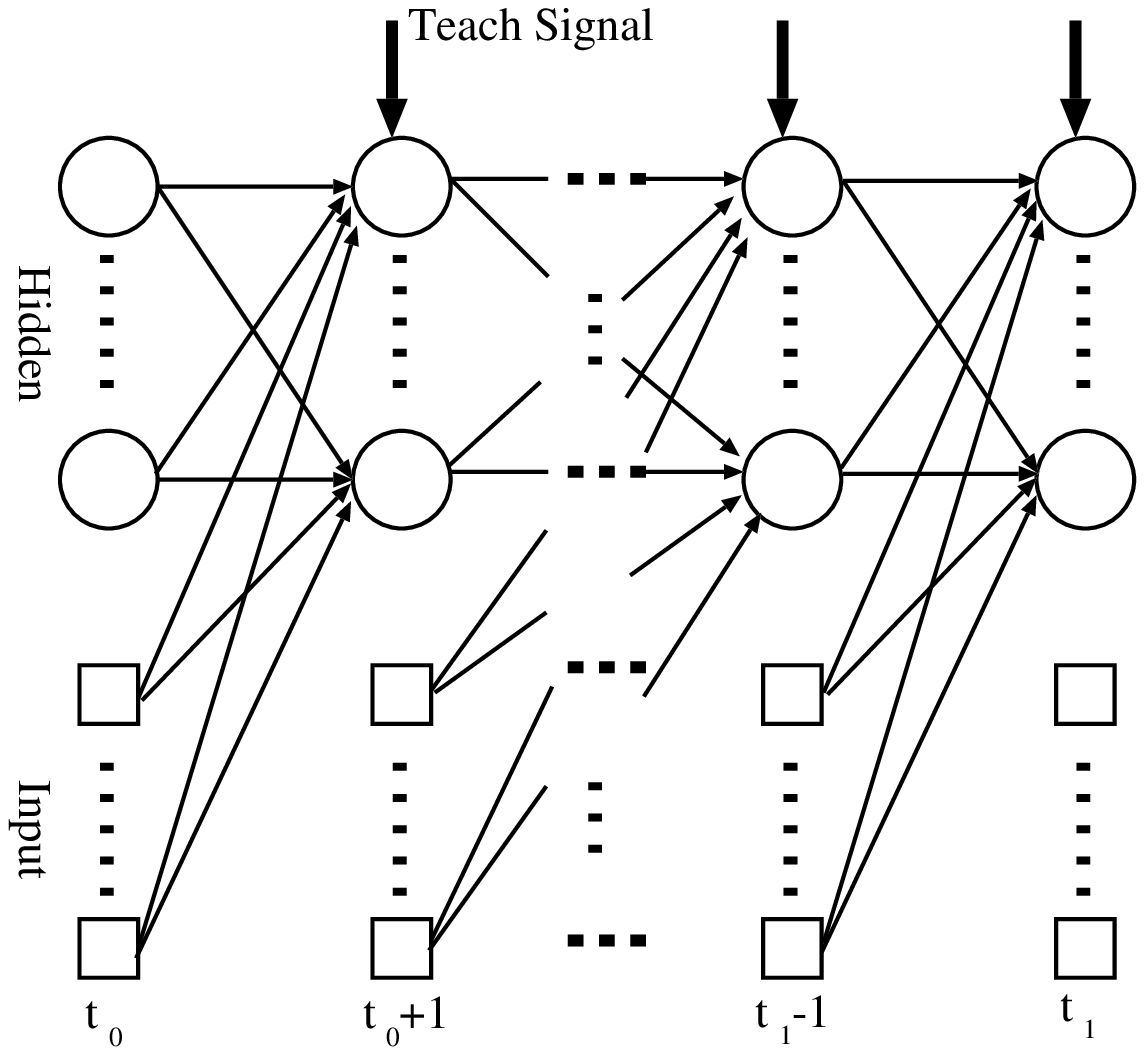}
\caption{Recurrent Neural Network}
\label{fig:RNN_EBPTT}
\end{center}
\end{figure}

Back Propagation through Time (BPTT) learning algorithm as shown in Fig.\ref{fig:RNN_EBPTT} is a variant of a BP algorithm for feedforward neural networks. In this algorithm, the update of the activation state of all units occurs in punctual moments of time. The first step involves the unfolding process of the network, which consists in replicating $t$ times the RNN obtaining an equivalent feedforward network. In this process each replicated connection shares its value $w_{ij}$ in all times. The resultant feedforward network can be trained by using the BP algorithm\cite{Grau2013}.

In the forward process of BP algorithm, the calculation of output $y_{i}$ at time $t$ of each neuron can be defined as 

\begin{eqnarray}
y_{i}(t)&=&f(x_{i}(t)) \\
\label{eq:rnn_01}
\nonumber x_{i}(t)&=&\sum_{j \in H} y_{j}(t)w_{ij} +\sum_{j \in I} x_{j}^{in}w_{ij}+\sum_{j \in M} y_{j}(t-\tau_{ij}) w_{ij}\\
\label{eq:rnn_02}
\end{eqnarray}
where $f$ refers to the neuron activation function, $H$ and $I$ denotes hidden layers indexes and the input neurons indexes, respectively. $x_{j}^{in}$ is the $j^{th}$ input neuron. $M$ is the index of neurons which store information about the previous network stage and $\tau_{ij} \ge 0$ is an integer value indicating the displacement in recurrent connections through times.

Then in the error BP process, each neuron $j$ is characterized by an error $\delta_{i}$. For the output layer, the error is defined as the difference between the expected value and the calculated value(\ref{eq:rnn_03}). For the hidden layers, it can be obtained taking into account the error in the successor layers(\ref{eq:rnn_04}).

\begin{eqnarray}
\delta_{j}(t)&=&(d_{j}-y_{j})y_{j}(1-y_{j}) \\
\label{eq:rnn_03}
\delta_{j}(t)&=&(y_{j}(1-y_{j})\sum_{i \in Suc(j)} w_{ij}\delta_{i} 
\label{eq:rnn_04}
\end{eqnarray}

The weight is updated by Eq.(\ref{eq:rnn_05}).
\begin{equation}
\Delta w_{ij}^{e+1} = \alpha \delta_{j} y_{j}
\label{eq:rnn_05}
\end{equation}
where $e$ refers to the order of the weight updates in the learning process. $\alpha$ is the learning rate.

\begin{figure}[btp]
\begin{center}
\includegraphics[scale=0.12]{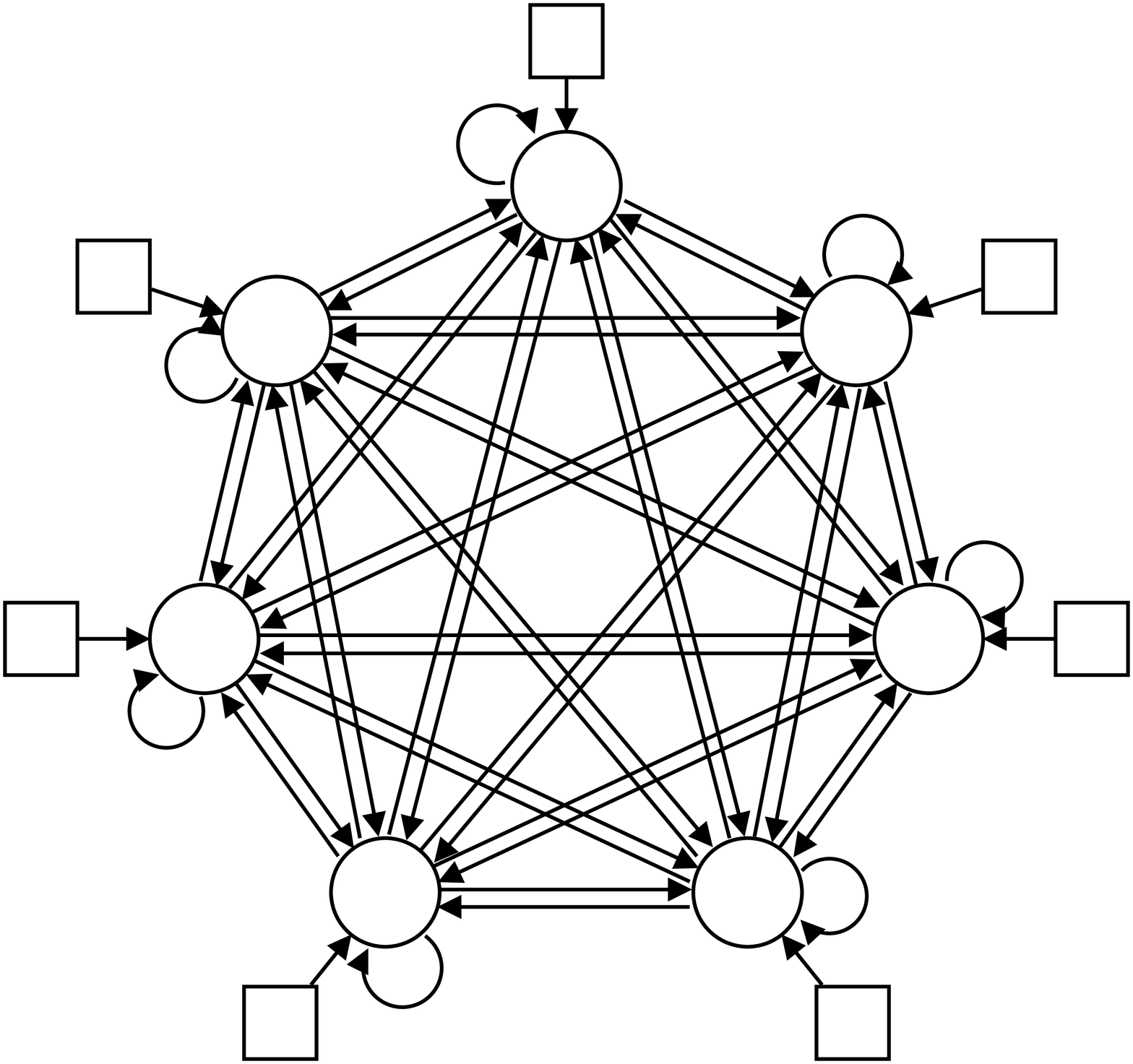}
\caption{Recurrent Neural Network for MSTN}
\label{fig:RNN_MSTN}
\end{center}
\end{figure}

Fig. \ref{fig:RNN_MSTN} shows the recurrent neural network for MSTN. The rectangles in Fig.\ref{fig:RNN_MSTN} are the input by the results of EGC and the circles are the hidden units. The weights are calculated from transition cost.

\subsection{Frequency of Output Patterns in BPTT-RNN}
\label{sec:FOP_BPTT-RNN}
The summation of transition costs from one mood to others in MSTN must be one, because the transition cost means the transition probability of moods. For this purpose, we calculated the output value to possible input patterns and checked the frequency of the pattern for the relation between input and output. Then the obtained frequency was set to the next transition cost. When 2 or more input nodes were fired simultaneously, each input node was set to the values which was divided 1 by the number of fired nodes.

\section{Experimental Results}
\label{sec:experimentalresults}

\subsection{Simulation Results for Movie Scenario}
\label{sec:moviescenario}
 In order to verify the effectiveness of the learning method, some learning results as shown in Table \ref{tab:RNN_OutputPattern1}-\ref{tab:RNN_OutputPattern7} were obtained. The 14 sequential episodes between a loving couple of 2 university students in the movie scenarios \cite{Ichikawa2003} were analyzed by the emotional orientated interface (EGC and MSTN) with learning algorithm (PS and BPTT-RNN). The bold values in each Table are trained by PS and BPTT-RNN. The values of the transition costs in Table \ref{tab:RNN_OutputPattern1}-\ref{tab:RNN_OutputPattern7} were converted into $[0,1]$ by the process in subsection \ref{sec:FOP_BPTT-RNN}.

\begin{table}[!htb]
\begin{center}
\tiny
\caption{Frequency for Output Pattern in RNN (Scenario 1)}
\vspace{-2mm}
\label{tab:RNN_OutputPattern1}
\begin{tabular}{l|llllllll}\hline
Current&\multicolumn{7}{c}{Next Mental State}\\\cline{2-8}
Mental State& Surprise   & Happy & Sad & Angry & Disgust & Fear & Normal\\ \hline
Surprise  &0.0220&  0.6151&  0.2986&  0.0154&  0.0126&  0.0127&  0.0236\\
Happy  &0.0107&  0.7762&  0.1580&  0.0108&  0.0100&  0.0093&  0.0248\\
Sad &0.0120&  {\bf 0.9111}&  0.0262&  0.0143&  0.0123&  0.0116&  0.0125\\
Angry &0.0144&  0.5689&  0.3227&  0.0295&  0.0166&  0.0133&  0.0246\\
Disgust &0.0139&  0.5757&  0.3221&  0.0208&  0.0293&  0.0138&  0.0244\\
Fear &0.0153&  0.5655&  0.3401&  0.0166&  0.0150&  0.0240&  0.0234\\
Normal &0.0012&  0.0285&  {\bf 0.9649}&  0.0013&  0.0012&  0.0011&  0.0018\\\hline
\end{tabular}
\end{center}
\end{table}

\begin{table}[!htb]
\begin{center}
\tiny
\caption{Frequency for Output Pattern in RNN (Scenario 2)}
\vspace{-2mm}
\label{tab:RNN_OutputPattern2}
\begin{tabular}{l|llllllll}\hline
Current&\multicolumn{7}{c}{Next Mental State}\\\cline{2-8}
Mental State& Surprise   & Happy & Sad & Angry & Disgust & Fear & Normal\\ \hline
Surprise  &0.0199&  0.7507&  0.1603&  0.0199&  0.0193&  0.0187&  0.0111\\
Happy  &0.0163&  {\bf 0.9064}&  0.0171&  0.0164&  0.0162&  0.0160&  0.0116\\
Sad &0.0206&  0.8183&  0.0881&  0.0211&  0.0205&  0.0199&  0.0115\\
Angry &0.0193&  0.7397&  0.1715&  0.0205&  0.0193&  0.0186&  0.0111\\
Disgust &0.0194&  0.7427&  0.1680&  0.0201&  0.0201&  0.0187&  0.0111\\
Fear &0.0193&  0.7377&  0.1739&  0.0197&  0.0192&  0.0191&  0.0110\\
Normal &0.0155&  0.5239&  0.4601&  0.0157&  0.0153&  0.0149&  0.0086\\\hline
\end{tabular}
\end{center}
\end{table}

\begin{table}[!htb]
\begin{center}
\tiny
\caption{Frequency for Output Pattern in RNN (Scenario 3)}
\vspace{-2mm}
\label{tab:RNN_OutputPattern3}
\begin{tabular}{l|llllllll}\hline
Current&\multicolumn{7}{c}{Next Mental State}\\\cline{2-8}
Mental State& Surprise   & Happy & Sad & Angry & Disgust & Fear & Normal\\ \hline
Surprise  &0.0189&  0.8461&  0.0675&  0.0191&  0.0188&  0.0183&  0.0113\\
Happy  &0.0168&  {\bf 0.8981}&  0.0237&  0.0170&  0.0168&  0.0165&  0.0112\\
Sad &0.0191&  0.8567&  0.0560&  0.0193&  0.0190&  0.0185&  0.0114\\
Angry &0.0188&  0.8443&  0.0694&  0.0191&  0.0187&  0.0183&  0.0113\\
Disgust &0.0189&  0.8448&  0.0688&  0.0191&  0.0188&  0.0183&  0.0113\\
Fear &0.0188&  0.8439&  0.0699&  0.0191&  0.0187&  0.0183&  0.0113\\
Normal &0.0180&  0.7955&  0.1222&  0.0182&  0.0179&  0.0175&  0.0108\\\hline
\end{tabular}
\end{center}
\end{table}

\begin{table}[!htb]
\begin{center}
\tiny
\caption{Frequency for Output Pattern in RNN (Scenario 4)}
\vspace{-2mm}
\label{tab:RNN_OutputPattern4}
\begin{tabular}{l|llllllll}\hline
Current&\multicolumn{7}{c}{Next Mental State}\\\cline{2-8}
Mental State& Surprise   & Happy & Sad & Angry & Disgust & Fear & Normal\\ \hline
Surprise  &0.0177&  0.8793&  0.0389&  0.0179&  0.0176&  0.0173&  0.0112\\
Happy  &0.0164&  {\bf 0.8989}&  0.0250&  0.0165&  0.0163&  0.0160&  0.0110\\
Sad &0.0177&  0.8804&  0.0377&  0.0179&  0.0177&  0.0173&  0.0112\\
Angry &0.0177&  0.8791&  0.0391&  0.0179&  0.0176&  0.0173&  0.0112\\
Disgust &0.0177&  0.8792&  0.0391&  0.0179&  0.0176&  0.0173&  0.0112\\
Fear &0.0177&  0.8791&  0.0392&  0.0179&  0.0176&  0.0173&  0.0112\\
Normal &0.0176&  0.8735&  0.0452&  0.0178&  0.0175&  0.0172&  0.0112\\\hline
\end{tabular}
\end{center}
\end{table}

\begin{table}[!htb]
\begin{center}
\tiny
\caption{Frequency for Output Pattern in RNN (Scenario 5)}
\vspace{-2mm}
\label{tab:RNN_OutputPattern5}
\begin{tabular}{l|llllllll}\hline
Current&\multicolumn{7}{c}{Next Mental State}\\\cline{2-8}
Mental State& Surprise   & Happy & Sad & Angry & Disgust & Fear & Normal\\ \hline
Surprise  &0.0170&  0.8904&  0.0310&  0.0171&  0.0169&  0.0166&  0.0111\\
Happy  &0.0162&   {\bf 0.8990}&  0.0256&  0.0163&  0.0161&  0.0159&  0.0109\\
Sad &0.0170&  0.8905&  0.0309&  0.0171&  0.0169&  0.0166&  0.0111\\
Angry &0.0170&  0.8904&  0.0310&  0.0171&  0.0169&  0.0166&  0.0111\\
Disgust &0.0170&  0.8904&  0.0310&  0.0171&  0.0169&  0.0166&  0.0111\\
Fear &0.0170&  0.8904&  0.0310&  0.0171&  0.0169&  0.0166&  0.0111\\
Normal &0.0169&  0.8899&  0.0315&  0.0171&  0.0169&  0.0166&  0.0111\\\hline
\end{tabular}
\end{center}
\end{table}

\begin{table}[!htb]
\begin{center}
\tiny
\caption{Frequency for Output Pattern in RNN (Scenario 6)}
\vspace{-2mm}
\label{tab:RNN_OutputPattern6}
\begin{tabular}{l|llllllll}\hline
Current&\multicolumn{7}{c}{Next Mental State}\\\cline{2-8}
Mental State& Surprise   & Happy & Sad & Angry & Disgust & Fear & Normal\\ \hline
Surprise  &0.0107&  0.4546&  0.3714&  0.0107&  0.0106&  0.0105&  0.1315\\
Happy  &0.0094&  0.0195&  0.0156&  0.0094&  0.0094&  0.0093& {\bf 0.9275}\\
Sad & 0.0106& {\bf 0.9174}&  0.0224&  0.0106&  0.0106&  0.0105&  0.0178\\
Angry &0.0107&  0.4546&  0.3714&  0.0107&  0.0106&  0.0105&  0.1314\\
Disgust &0.0107&  0.4546&  0.3713&  0.0107&  0.0106&  0.0105&  0.1315\\
Fear &0.0107&  0.4546&  0.3712&  0.0107&  0.0106&  0.0105&  0.1316\\
Normal &0.0106&  0.0190& {\bf 0.9201}&  0.0106&  0.0106&  0.0105&  0.0186\\\hline
\end{tabular}
\end{center}
\end{table}

\begin{table}[!htb]
\begin{center}
\tiny
\caption{Frequency for Output Pattern in RNN (Senario 7)}
\label{tab:RNN_OutputPattern7}
\begin{tabular}{l|llllllll}\hline
Current&\multicolumn{7}{c}{Next Mental State}\\\cline{2-8}
Mental State& Surprise   & Hayppy & Sad & Angry & Disgust & Fear & Normal\\ \hline
Surprize  &0.1514&  0.2629&  0.4826&  0.0207&  0.0206&  0.0204&  0.0414\\
Happy  &{\bf 0.7919}&  0.0503&  0.0502&  0.0221&  0.0219&  0.0218&  0.0418\\
Sad &0.1313&  0.6018&  0.1876&  0.0179&  0.0178&  0.0177&  0.0259\\
Angry &0.1514&  0.2629&  0.4827&  0.0207&  0.0206&  0.0204&  0.0414\\
Disgust &0.1514&  0.2629&  0.4826&  0.0207&  0.0206&  0.0204&  0.0414\\
Fear &0.1514&  0.2630&  0.4825&  0.0207&  0.0206&  0.0204&  0.0414\\
Normal &0.1044&  0.0578&  0.7747&  0.0143&  0.0142&  0.0141&  0.0206\\\hline
\end{tabular}
\end{center}
\end{table}

\subsection{Personality Trait Based Mood in MSTN}
\label{sec:bigfivemakers}
The Big Five Makers(BFM)\cite{Goldberg1992} was designed to assess the constellation of traits defined by the Five Factor Theory of Personality as follows:
\begin{itemize}
\item Openness is characterized by originality, curiosity, and ingenuity. This factor is sometimes referred to as Culture because of its emphasis on intellectualism, polish, and independence of mind. This factor is also sometimes referred to as Intellect because of its emphasis on intelligence, sophistication, and reflection.
\item Conscientiousness is characterized by orderliness, responsibility, and dependability. This factor is sometimes referred to as Dependability.
\item Extraversion is characterized by talkativeness, assertiveness, and energy. This factor is sometimes referred to as Surgency.
\item Agreeableness is characterized by good-naturedness, cooperativeness, and trust.
While this factor is most commonly called Agreeableness, it can also be seen as a combination of friendliness and compliance.
\item Neuroticism is characterized by upsetability and is the polar opposite of emotional stability. This factor is sometimes scored in the opposite direction and referred to as Emotional Stability.
\end{itemize}

In the section, we define the relation among the terms in BFM and the the current mood to next mood in MSTN as shown in Table \ref{tab:MSTN_PersonaliryTrait}.

\begin{table}[!htb]
\scriptsize
\centering
\caption{The relation between MSTN and BFM}
\label{tab:MSTN_PersonaliryTrait}
\begin{tabular}{c|c|c}\hline
current & next & Personality Trait \\ \hline
\multirow{7}{*}{Surprise}
&Surprise&Neuroticism\\
&Happy&Conscientiousness, Neuroticism\\
&Sad&Agreeableness\\
&Angry&{\bf Extraversion}, {\bf Conscientiousness}\\
&Disgust&{\bf Conscientiousness}\\
&Fear&Agreeableness\\
&Normal&{\bf Openness}, Extraversion\\\hline
\multirow{7}{*}{Happy}
&Surprise&Neuroticism\\
&Happy&Openness, Conscientiousness, Neuroticism\\
&Sad&Agreeableness\\
&Angry&{\bf Extraversion}, {\bf Conscientiousness}\\
&Disgust&{\bf Conscientiousness}\\
&Fear&Agreeableness\\
&Normal&{\bf Openness}, Extraversion\\\hline
\multirow{7}{*}{Sad}
&Surprise&-\\
&Happy&Agreeableness\\
&Sad&Agreeableness, {\bf Extraversion}\\
&Angry&{\bf Extraversion}, {\bf Conscientiousness}\\
&Disgust&{\bf Conscientiousness}\\
&Fear&Agreeableness\\
&Normal&{\bf Agreeableness}, {\bf Openness}, Extraversion\\\hline
\multirow{7}{*}{Anger}
&Surprise&-\\
&Happy&-\\
&Sad&Agreeableness\\
&Angry&{\bf Extraversion}, {\bf Conscientiousness}\\
&Disgust&{\bf Conscientiousness}\\
&Fear&Agreeableness\\
&Normal&{\bf Openness}, Extraversion\\\hline
\multirow{7}{*}{Fear}
&Surprise&-\\
&Happy&-\\
&Sad&Agreeableness\\
&Angry&{\bf Extraversion}, {\bf Conscientiousness}\\
&Disgust&{\bf Extraversion}, {\bf Conscientiousness}\\
&Fear&Agreeableness\\
&Normal&{\bf Openness}, Extraversion\\\hline
\multirow{7}{*}{Fear}
&Surprise&-\\
&Happy&-\\
&Sad&Agreeableness\\
&Anger&{\bf Extraversion}, {\bf Conscientiousness}\\
&Disgust&{\bf Conscientiousness}\\
&Fear&Agreeableness\\
&Normal&{\bf Openness}, Extraversion\\\hline
\multirow{7}{*}{Normal}
&Surprise&Neuroticism\\
&Happy&Openness, Conscientiousness, Neuroticism\\
&Sad&Agreeableness\\
&Anger&{\bf Extraversion}, {\bf Conscientiousness}\\
&Disgust&{\bf Conscientiousness}\\
&Fear&Agreeableness\\
&Normal&{\bf Openness}, Extraversion, {\bf Conscientiousness}\\\hline
\multicolumn{3}{l}{Note: The term in Bold the inverse of the original meaning.}
\end{tabular}
\end{table}

\subsection{Personality Trait from the Transition Costs}
 In Scenario 1, the mental detour (`Normal' $\rightarrow$ `Sad' $\rightarrow$ 'Happy') was obtained by PS method and Table \ref{tab:RNN_OutputPattern1} shows that the transition costs of (`Normal' $\rightarrow$ `Sad') and (`Sad' $\rightarrow$ 'Happy') were higher than before training. Moreover, the transition costs in the other paths to 'Sad' and to 'Happy' except the detour path were higher. Such observations were also seen under the environment of all scenarios in the experimental results.

\section{Conclusion}
In this paper, the transition costs are trained by Profit Sharing and recurrent neural network learning algorithms to express the user's personality trait based mood. The transition of mental states in MSTN can represent the Five Factor of personality. We discussed the effectiveness from the simulation results by using some episodes of movie scenario. We will develop the emotional orientated interface in web application in future.

\section*{Acknowledgment}
This work was supported by JSPS KAKENHI Grant Number 25330366.

\end{document}